\documentclass[aps,twocolumn,showpacs,amsmath]{revtex4}
\usepackage{graphicx}
\usepackage{slashbox}

\begin{document}

\title{Observing Collapse in Colliding Two Dipolar Bose-Einstein Condensates}

\author{B. Sun and M. S. Pindzola}
\affiliation{Department of Physics, Auburn University, Auburn,
Alabama, 36849, USA}

\begin{abstract}
We study the collision of two Bose-Einstein condensates with pure
dipolar interaction. A stationary pure dipolar condensate is known
to be stable when the atom number is below a critical value.
However, collapse can occur during the collision between two
condensates due to local density fluctuations even if the total
atom number is only a fraction of the critical value. Using full
three-dimensional numerical simulations, we observe the collapse
induced by local density fluctuations. For the purpose of future
experiments, we present the time dependence of the density
distribution, energy per particle, and the maximal density of the
condensate. We also discuss the collapse time as function of the
relative phase between the two condensates.
\end{abstract}


\maketitle

\section{Introduction}
Since the experimental observation of $^{52}$Cr Bose-Einstein
condensate (BEC) \cite{pfau1}, there has been a growing interest
in the study of ultracold dipolar gases. $^{52}$Cr has a magnetic
dipole moment of $\mu=6\mu_B$ ($\mu_B$ is the Bohr magneton) which
has at least 36 times larger dipolar interaction strength than its
alkaline counterparts. Therefore, $^{52}$Cr is an ideal choice for
investigating novel dipolar effects in neutral atoms. It has been
shown theoretically and experimentally that there are detectable
modifications to the condensate density profile
\cite{pfau0,pfaua,pfau3,meystre,pfau2,bohn1,pfau4,pfau00} and
elementary excitations \cite{you1,santos1,eberlin,bohn3,pfau5} due
to this long ranged and anisotropic interaction.

The stability of a dipolar condensate is a fundamental question to
answer since the dipolar interaction is partially attractive. One
feature of the dipolar condensate is that the effective dipolar
interaction depends on the shape of the trap. This can be roughly
understood from a simple argument based on the competition between
the potential energy per particle $u_p$ and the dipolar
interaction energy per particle $u_d$. Suppose the dipolar gas is
polarized along the $z$-axis and confined in a cylindrically
symmetric trap with an aspect ratio
$\lambda=\omega_z/\omega_{\perp}$. Without the trapping potential,
the dipoles tend to arrange into a head-to-tail configuration and
lowers $u_d$ which results in an unstable condensate. This is also
true for a prolate trap ($\lambda\ll 1$) because this
configuration also lowers $u_p$. However, in an oblate trap
($\lambda\gg 1$), there is a competition between $u_p$ and $u_d$.
Although $u_p$ is almost independent of the atom number $N$, the
magnitude of $u_d$ increases as $N$ increases in general. When
$u_p$ dominates $u_d$, i.e. $N$ is below a critical value $N_c$,
the dipoles are prone to arrange into a head-to-head configuration
which is stable. In the opposite case, the gas still favors a
head-to-tail configuration which is again unstable.

The dependence of this stability on the trap aspect ratio and atom
number is investigated more thoroughly in a recent publication
where the authors show that the stability diagram exhibits richer
physics beyond our intuitive understanding \cite{bohn1}. As shown
in the stability diagram from Ref. \cite{bohn1}, an increase in
$\lambda$ will tend to stabilize a dipolar condensate and for a
given $\lambda$, there is always a critical value of $N_c$ above
which the condensate is unstable, in agreement with our simple
analysis. However, the dipolar interaction can cause the formation
of a structured condensate, e.g. a biconcave one, in addition to a
normal Gaussian shaped condensate. The underlying mechanisms of
the instability can be analyzed from the Bogoliubov-de Gennes
equation which are referred to as angular- and radial-roton
instability, respectively \cite{santos1,bohn1}.

Previous studies concerning collapse in dipolar gases have focused
on the response of a stationary condensate to a modulation of the
s-wave scattering length \cite{pfau4,dell00,dell01,bohn2r}. In
this paper, we want to investigate the possibility of observing
collapse induced by purely dipolar interaction in a dynamic
collision process. For this purpose, we will study the collision
dynamics of two dipolar condensates and discuss the collapse
effect. A similar scenario of overlapping several independent
condensates is briefly discussed where the effect of collapse on
the interference fringes is observed \cite{pfau00}. Another
relevant scenario is for the case of pure attractive s-wave
scattering where colliding two bright solitary waves may also lead
to collapse \cite{adams1}. We will give a more thorough study and
present more detailed results such as density distributions and
the collapse time to support future experiments. The structure of
this paper is as follows. In Sec. II, we start with the
generalized Gross-Pitaevskii equation, giving our simulation
parameters and the numerical scheme. In Sec. III, we present
numerical results and talk about the collapse effect in detail.
Finally, we give the conclusion in Sec. IV.

\section{Theory}

The dynamics of the two BECs at sufficiently low temperature are
described by the generalized Gross-Pitaevskii equation (GPE),
\begin{eqnarray}
i\hbar {\partial \psi({\bf r},t) \over \partial t}&=&
(H_0+H_s+H_d)\psi({\bf r},t) \label{ht}
\end{eqnarray}
with various terms listed below
\begin{eqnarray}
H_0&=& -{\hbar^2\over 2m}\nabla^2 +V_{\rm trap}({\bf r},t)
\nonumber\\
H_s&=&N{4\pi \hbar^2 a_0\over m} |\psi({\bf r},t)|^2 \nonumber\\
H_d&=& N \int d{\bf r}' V_{\rm dd}({\bf r }-{\bf r }')|\psi({\bf
r}',t)|^2, \nonumber
\end{eqnarray}
where the wave function is always normalized to unity. $m$ is the
atom mass, $N$ is the atom number, and $a_0$ is the $s$-wave
scattering length. The trap potential assumes the following form
\begin{eqnarray}
V_{\rm trap}({\bf r},t) &=& {1\over 2}m  \omega_{\perp}^2
(\rho^2+\lambda^2 z^2) + {\cal A} e^{-{x^2\over 2w^2}} \theta(-t),
\nonumber
\end{eqnarray}
where ${\boldsymbol \rho}=(x,y)$ and $\theta(\cdot)$ is the
Heaviside step function. It describes a trap potential which at
$t=0$ is a combination of a cylindrically harmonic trap
($\lambda=\omega_z/\omega_{\perp}$) plus a central Gaussian
barrier (with height ${\cal A}$ and width $w$) along the $x$-axis
and the barrier is removed immediately after $t=0$. The dipolar
interaction potential for a gas polarized along the $z$-axis is
\begin{equation}
V_{\rm dd}({\bf r })= c_d{r^2-3z^2 \over r^5}.\nonumber
\end{equation}
$c_d=\mu_0 \mu^2/(4\pi)$ where $\mu_0=4\pi\times 10^{-7}$
T$\cdot$m/A is the vacuum permeability and $\mu$ is the magnetic
dipole moment of the atom.


We adopt the length scale $a_{ho}=\sqrt{\hbar/(m\omega_{\perp})}$
and the time scale $1/\omega_{\perp}$ as those of a harmonic
oscillator and substitute ${\bf r}\to {\bf r}/a_{ho}$, $t\to t
\omega_{\perp}$, ${\cal A}\to {\cal A}/(\hbar\omega_{\perp})$,
$w\to w/a_{ho}$, $Q=N a_0/a_{ho}$, and $D=N
c_d/(\hbar\omega_{\perp}a_{ho}^3)$ into Eq. (\ref{ht}). We then
arrive at the dimensionless version of the generalized GPE
\begin{eqnarray}
i{\partial \psi({\bf r},t) \over \partial t}=\left(-{1\over
2}\nabla^2+\tilde{V}_{\rm trap}({\bf r},t)+ 4\pi Q |\psi({\bf
r},t)|^2
\right.\nonumber\\
\left.+ D \int d{\bf r}' {|{\bf r}-{\bf r}'|^5-3(z-z')^2 \over
|{\bf r}-{\bf r}'| ^5}|\psi({\bf r}',t)|^2 \right) \psi({\bf
r},t), \label{htt}
\end{eqnarray}
where $\tilde{V}_{\rm trap}({\bf r},t)={1\over 2}(\rho^2+\lambda^2
z^2)+{\cal A} e^{-{x^2\over 2w^2}}\theta(-t)$. In this paper, we
study the dynamics of BECs with purely dipolar interaction,
assuming the $s$-wave scattering length can be tuned to zero
($Q=0$) by modulating the magnetic field. The interaction term
(the last term in Eq. (\ref{htt})) can be conveniently computed by
making use of the Fast Fourier transform ${\cal F}$ with the help
of the following identities \cite{pfau0}
\begin{eqnarray}
{\cal F}[H_d]=N{\cal F}[V_{\rm dd}({\bf r})]{\cal F}[|\psi({\bf r},t)|^2],\nonumber\\
{\cal F}[V_{\rm dd}({\bf r})]=c_d{4\pi\over 3}(3\cos^2\theta_{\bf
k}-1),
\end{eqnarray}
where $\theta_{\bf k}$ is the angle between the conjugate momentum
${\bf k}$ and the $z$-axis. The $H_d$ term can be obtained via an
inverse Fast Fourier transform.

\section{Numerical Results}

In this section, we present our numerical results for colliding
two BECs. We first briefly discuss the initial state. Then we will
investigate the collision dynamics in detail.

\begin{figure}[htb]
\includegraphics[width=3.25in]{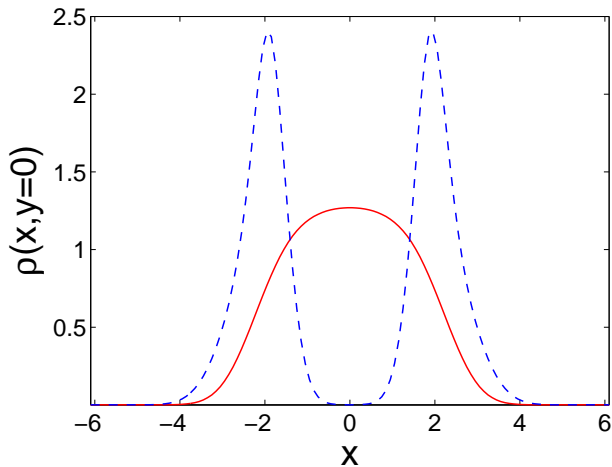}
\caption{Column density $\rho(x,y)$ along $y=0$ as function of
$x$. The red solid (blue dashed) curve is for ground state in the
single harmonic trap without (with) a central gaussian barrier.}
\label{figure1}
\end{figure}

We use $^{52}$Cr atom with $\mu=6\mu_B$ in our numerical
simulations. For the harmonic trap, the transverse frequency is
fixed at $\omega_{\perp}=(2\pi)25$ Hz. The corresponding length
scale is $a_{ho}=2.78\mu $m and the time scale is
$1/\omega_{\perp}=6.37$ms. The mesh in all three directions is
$192\times 0.1$. The initial state at $t=0$ is obtained first by
imaginary time relaxation. We then add a phase factor $e^{i\phi}$
to the right side of the wave function to account for a
possibility of uncertainty in the initial phases between the two
BECs \cite{ketterle2}. To verify the reliability of our numerical
results, we search for the critical atom numbers for the ground
state in harmonic traps with different aspect ratios and compare
them with those reported in Ref. \cite{bohn1}. We find our results
for the ground state are in good agreement with Ref. \cite{bohn1}.
For example, we found for $\lambda=2$ ($\omega_z=(2\pi)50$Hz), the
critical atom number for the ground state is around $N_c=4400$. We
also search the critical atom numbers for other excited states.
For example, for $p$-wave soliton, the critical atom number is
found to be larger than the ground state for the same aspect
ratio. Here the $p$-wave soliton refers to the state with a node
along one direction (say $x$) so the ansatz function for the
imaginary time propagation takes the form of $x
e^{-\rho^2/2-\lambda z^2/2}$. The detailed study of excited states
will be reported elsewhere. In this paper, we are interested in
the situation of colliding two ground state condensates with a
strong confinement along the $z$-axis so we choose $\lambda=10$
($\omega_z=(2\pi)250$Hz) and $N=19000$ where the ground state is
found to be stable. We will use them throughout our simulation for
the collision dynamics. The column density $\rho(x,y)$ along $y=0$
is shown as the red solid curve in Fig. \ref{figure1} where
$\rho(x,y)=\int dz |\psi({\bf r})|^2$. When a central barrier is
added, it is shown as the blue dashed curve in Fig. \ref{figure1}.
The dipolar interaction strength is computed to be $D=16.5$. The
parameters for the central barrier are ${\cal A}=15$ and $w=1$
which are chosen so that, at the initial time $t=0$, there is no
substantial overlap between the two condensates. We want to
emphasize that this atom number is far below the critical value
$N_c\simeq 70000$ estimated from \cite{bohn1} for the same single
harmonic trap, as the motivation is to observe the local collapse
induced by density fluctuations rather than the global collapse
induced by an overall critical atom number.  After $t>0$, the
condensates are out of equilibrium and start to collide with each
other and oscillate in the trap. We explore the subsequent
dynamics using real time propagation.

\begin{figure}[htb]
\includegraphics[width=3.25in]{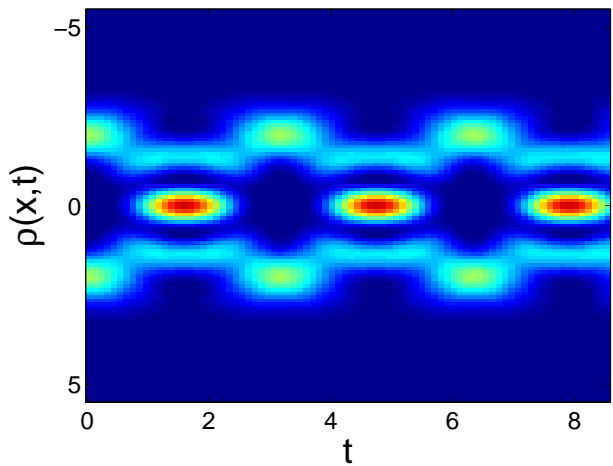}
\caption{Linear density distribution $\rho(x)$ as function of time
$t$ for two non-interacting condensates with $\phi=0$. Blue (red)
stands for the minimum (maximum) density.} \label{figure2}
\end{figure}

We first discuss the simplest case of two non-interacting BECs
($Q=0$ and $D=0$), assuming the dipolar interaction can be tuned
to zero \cite{pfau6}. In Fig. \ref{figure2}, we show our numerical
results of linear density $\rho(x,t)$ as function of time $t$ for
the relative phase $\phi=0$. $\rho(x,t)$ is obtained by
integrating along both the $y$- and $z$-axis, i.e.
$\rho(x,t)=\int\int dydz |\psi({\bf r},t)|^2$. At the initial
time, the two BECs are well separated. After the barrier is turned
off, the two BECs start to collide with each other. As a
consequence, we can see an interference pattern between them. The
interference pattern disappears when the two BECs pass through
each other. The two BECs are then reflected by the harmonic trap
and ready for the collision in the next cycle. Such a necklace
pattern of $\rho(x,t)$ is expected to keep repeating itself, with
only a few cycles are selectively shown in Fig. \ref{figure2}.
Because the two BECs are non-interacting, this pattern persists
and collapse will not take place.

\begin{figure}[htb]
\includegraphics[width=3.25in,angle=270]{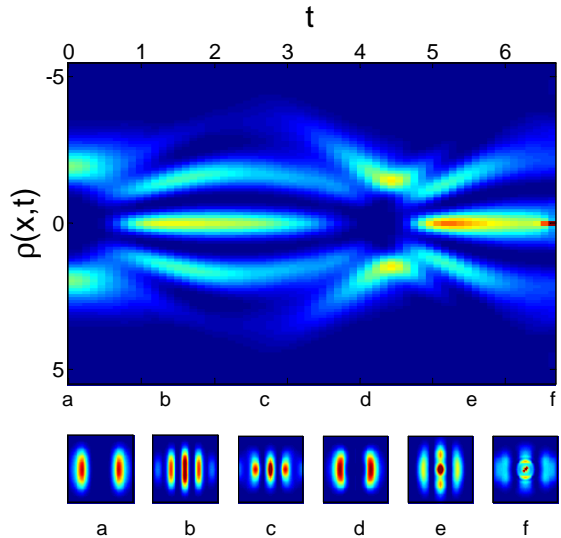}
\caption{Upper figure: linear density distribution $\rho(x)$ as
function of time $t$ for $\phi=0$. Lower figure: column density
$\rho(x,y,t)$ as 6 different times labelled by (a)-(f). Field of
view in each subplot is $(x,y)=[-3.5,3.5]\times [-3.5,3.5]$. In
both figures, blue (red) stands for the minimum (maximum) density.
However, the colormap of the upper figure is different from that
of the lower figure and is adjusted for better visual effect. }
\label{figure3}
\end{figure}

\begin{figure}[htb]
\includegraphics[width=3.25in]{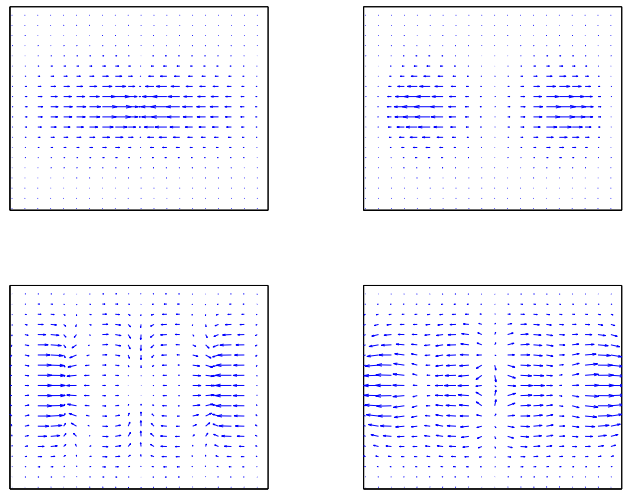}
\caption{Probability current $J$ for $\phi=0$ at $t=1.43$ (left)
and $t=2.86$ (right). The upper (low) row is for the
non-interacting (dipolar) case. Field of view in each subplot is
$(x,y)=[-3,3]\times [-3,3]$.} \label{figure4}
\end{figure}

\begin{figure}[htb]
\includegraphics[width=3.00in]{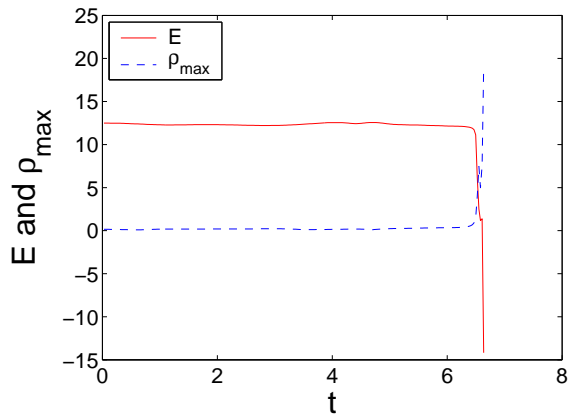}
\caption{Energy per particle $E$ (red solid) and the maximal
density $\rho_{\rm max}$ (blue dashed) as function of time $t$
$\phi=0$.} \label{figure5}
\end{figure}

Now we present the numerical results for dipolar condensates
($Q=0$ and $D=16.5$). Firstly, the density distribution differs
significantly from that in the non-interacting case. Secondly and
more importantly, we observe the collapse of the two BECs which is
absent in the non-interacting case. Our numerical results for
relative phase $\phi=0$ are presented in Figs.
\ref{figure3}-\ref{figure5}. In Fig. \ref{figure3}, we show the
density distributions $\rho(x,t)$ and $\rho(x,y,t)$ as function of
time $t$. From the upper figure in Fig. \ref{figure3}, we can see
that the two dipolar BECs behave similarly to non-interacting
BECs: they collide with each other, interfere, and are reflected
by the harmonic trap. However, the differences between them still
merit some discussions here. One difference is that the period of
the first cycle is about 4.5 which is larger than that of
non-interacting BECs ($\sim 3$). This slowed motion is reminiscent
of the damping effect for non-dipolar condensates ($Q\neq 0$ and
$D=0$) where the interaction causes complex flow patterns acting
as a damping force \cite{scott,bo1,bo2}. Here we also find complex
flow patterns for the dipolar interaction. Selected plots for the
probability current $\vec J ={\hbar\over m}{\rm
Im}[\psi^*(x,y,z,t)\nabla\psi(x,y,z,t)]$ ($`` \rm{Im}"$ denotes
the imaginary part) are shown in Fig. \ref{figure4}. The upper row
of Fig. \ref{figure4} is for non-interacting condensates where the
flow lines are all along the $x$-axis. However, in the lower row
of Fig. \ref{figure4} as for interacting condensates, we can see
flow lines bent by the interaction. As a result of the damped
motion, roughly 5 interference fringes can be seen, cf only 3
interference fringes can be seen for the non-interacting case.
Another difference is the increased density in the central region
at the final time ($t\sim 6.5$) which leads to the collapse
effect. Since the total atom number is much lower than the
critical atom number, this collapse is purely seeded by local
density fluctuations. In this case, the interference is
responsible for the enhancement in the density distribution.
Although the dipolar interaction is cylindrically symmetric, the
existence of the initial central barrier breaks this symmetry. As
a result, the density distribution will develop an anisotropic
pattern in the transverse plane. This cannot be seen from the
linear density but instead can be seen from the column density
$\rho(x,y,t)$. In the lower part of Fig. \ref{figure3}, we show 6
time snapshots (a-f) for $\rho(x,y,t)$ whose times are also marked
on the horizontal axis of the upper figure. The density patterns
in Figs. 3(a)-(d) clearly show a cycle of collision. We can see
that, due to the dipolar interaction, the density pattern of Fig.
3(d) is not the same as that of Fig. 3(a). At a later time $t\sim
5.6$ (Fig. 3(e)), the density also exhibits a distribution along
the $y$-axis: the density is maximal in the central region and
surrounded by 4 lobes. At the final time $t\sim 6.5$ (Fig. 3(f))
just before the collapse, the central density maximum evolves into
a singularity and triggers the collapse. Note that this
singularity is purely artificial since GPE cannot handle the
post-collapse dynamics. However, the GPE can still provide an
accurate prediction of the onset of this collapse. Therefore, we
still choose to present Fig. 3(f) to demonstrate the singular
density profile. From the definition of $D$, where $N/a_{ho}^3$ is
just the averaged density, it is perhaps not surprising that the
local density fluctuations may induce collapse even though the
total atom number is far below the critical atom number. What is
somewhat more interesting is that the collapse does not happen
during the first cycle of collision ($t\sim 0-4.5$) but only at a
later time ($t \sim 6.5$). This is different from the previous
study for the case of attractive s-wave scattering where the
collapse happens during the first cycle of collision
\cite{adams1}. In our case, it takes longer time for the
anisotropic dipolar interaction to build up the local density
fluctuation and eventually singularity.

To track this collapse more quantitively, we show the energy per
particle $E$ and the maximal density $\rho_{\rm max}$ as function
of time $t$ in Fig. \ref{figure5}. $\rho_{\rm max}\equiv {\rm
max}_{\{\bf r \}}|\psi({\bf r})|^2$ at each time step. The heating
effect arising from the sudden removal of the barrier potential is
negligible as the energy variation at the initial stage is about
$0.5\hbar\omega_{\perp}/k_B\sim$  nK, so the description using GPE
is still a good approximation. It is clear from Fig. \ref{figure5}
that the collapse is signaled by a sudden drop (increase) in $E$
($\rho_{\rm max}$) due to the singularity developed in the
condensate density. Here we are only interested in the onset of
this collapse. Although the collapse dynamics afterwards are also
interesting, a similar scenario of collapse has already been
experimentally observed and discussed in detail in Ref.
\cite{pfau4} and will not be discussed in this paper.

\begin{figure}[htb]
\includegraphics[width=2.75in]{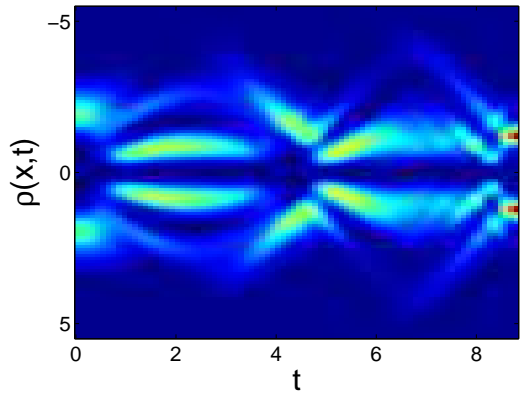}
\caption{Linear density distribution $\rho(x)$ as function of time
$t$ for $\phi=\pi$. Blue (red) stands for the minimum (maximum)
density.} \label{figure6}
\end{figure}

\begin{figure}[htb]
\includegraphics[width=3.00in]{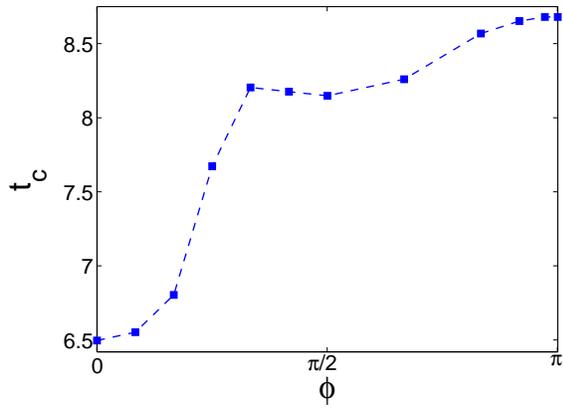}
\caption{Collapse time $t_c$ as function of the relative phase
$\phi$. Data points with blue square are numerical results. Dashed
line is a guide to eyes.} \label{figure7}
\end{figure}

For other choices of $\phi$, we also observe the collapse effect.
For example, in Fig. \ref{figure6}, we show $\rho(x,t)$ as
function of time $t$ for $\phi=\pi$. We find that the period of
the first cycle is close to that of $\phi=0$. Since the
condensates are always in a phase of $\pi$, the two BECs never
pass through each other. It seems as if they just collide and are
bounced back from each other. However, we still observe strong
density fluctuations and the induced collapse at $t\sim 8.7$. In
this case, the density maximum is not in the center, rather it is
found at two different locations at the same time. The complete
dependence of the collapse time $t_c$ on $\phi$ is shown in Fig.
\ref{figure7}. $t_c$ is defined as the time when
$|1-E(t_c)t_c/\int_0^{t_c} E(t) dt|=\delta$ and we choose
$\delta=5\%$ here. Although a purely empirical definition, $t_c$
does capture the onset of the energy variation due to the
diverging density profile. We find that a moderate change in
$\delta$ does not change our conclusion qualitatively. The overall
trend of the curve is that $t_c$ increases as $\phi$ increases.
This can be explained as follows. For small $\phi \sim 0 $, the
collision behaves as ``attractive" and the maximal density is
likely to be found in the trap center resulting in a enhanced
density. While for large $\phi \sim \pi$, the collision behaves as
``repulsive" and the density maximum is likely to be located off
the trap center resulting in a reduced density. In other words, it
takes longer time to accumulate high enough density for larger
$\phi$.

\section{conclusion}
In conclusion, we have studied the collision of two Bose-Einstein
condensates with pure dipolar interaction. A stationary dipolar
condensate is known to be stable when the atom number is below a
critical value. However, we find that, even though the total atom
number is just a fraction of the critical value, during the
collision of two condensates, the local density fluctuations can
still induce the collapse. To demonstrate this, we have performed
full three-dimensional numerical simulations for typical
experimental parameters. We present density distributions as
function of time for different relative phases and compare them
with those of two non-interacting condensates. We find that the
dipolar interaction modifies the density profiles significantly.
In addition, we show the collapse time as function of the relative
phase between the two condensates. It turns out that a larger
relative phase tends to increase the collapse time. We hope our
study can be helpful to the ongoing experiments with degenerate
dipolar gases.

This work is supported in part by an NSF grant to Auburn
University. Computational work was carried out at the National
Energy Research Scientific Computing Center in Oakland,
California.

\end{document}